# Two-dimensional collisions and conservation of momentum

Lior M. Burko

Theiss Research, La Jolla, CA and Peachtree Ridge High School, Suwanee, GA

## Introduction

Analysis of collisions is standardly included in the introductory physics course. In one dimension (1D), there do not seem to be any unusual issues: Typically, the initial velocities of the two colliding objects are specified, and the problem is to find the final velocities. In 1D there are therefore two unknown variables. One can write the equation for conservation of momentum, and either the equation for conservation of energy for the perfectly elastic case, or the expression for the coefficient of restitution (COR) otherwise. Thus, one has two equations for two unknowns, and one may solve the problem fully. An issue arises, however, in two-dimensional (2D) collisions: There are four unknown variables (two components of the final velocity of each object), but now there appear to be only three equations: two components of the equation of conservation of momentum, and the energy condition. The problem may appear therefore to be underdetermined. If this problem were in principle an underdetermined one, one would fail in predicting the outcome of the collision experiment. We describe how one may assign students an appropriate lab exercise and problems for an interesting class of 2D collisions for which one can determine uniquely the outcome of the collision.

For detail on approaches taken by textbooks see the online Appendix [1]. Common to all textbooks we have surveyed is that in practice they solve problems that reduce the number of unknowns to three by either considering the case of vanishing COR, or to the case in which (partial) information on the final conditions of the system is given. These practical simplifications are very reasonable: One often considers particular cases for which one can set a well-posed problem that leads to the solution (e.g., in Rutherford scattering one typically measures the scattering angle). We present an approach that allows us to solve the problem for all four unknowns for an interesting class of problems.

The hallmark of physics is that we can predict the final condition of the system given a full set of initial conditions and the equations of motion. Taken at face value, statement that information on the final state must be had might mislead a reader into concluding that predictability breaks down in principle, and that even if a full set of initial conditions were given in addition to the equations of motion, there would still be no unique solution to the problem unless (partial) final conditions are also imposed. How does nature select then a specific solution from a family of solutions?

A more nuanced approach is that one can determine the future evolution of the system from its initial state either by using the equations of motion, or by using conserved quantities (or a combination thereof). Conservation laws constrain the parameter space of the solutions. (This approach is emphasized in [2].) In some cases, such as 1D collisions, the conservation laws are sufficient in order to uniquely determine the outcome. However, unless more information about the system than is included in conserved quantities is given, only a family of possible solutions is found instead of a unique solution. Such additional information can be supplied by the equations of motion, or by details of the interaction. In practice, one can also use (partial) information about the final outcome, as in the case of Rutherford scattering.

### A simplified model

Additional detail of the interaction can allow us to use conservation laws to obtain a unique solution for an interesting class of 2D collisions. Two-dimensional collisions have been discussed extensively from multiple viewpoints. Typically, the collision is treated either in a reference frame in which one of the colliding objects is at rest before the collision [3,4], in the laboratory frame [5], or in the center of mass frame [5]. Here, we use a special coordinate system in the laboratory frame.

For specificity, we use a video clip of a 2D collision [6], which shows two equal mass colliding pucks on an air table. Figure 1 shows snapshots from the video. The main characteristic of the class of problems we are interested in is that the pucks do not rotate. The problem the students are given is to predict the final velocities of the two pucks given the (normal) COR and the initial velocities. Then, the students measure the final velocities of the pucks and compare with their predictions.

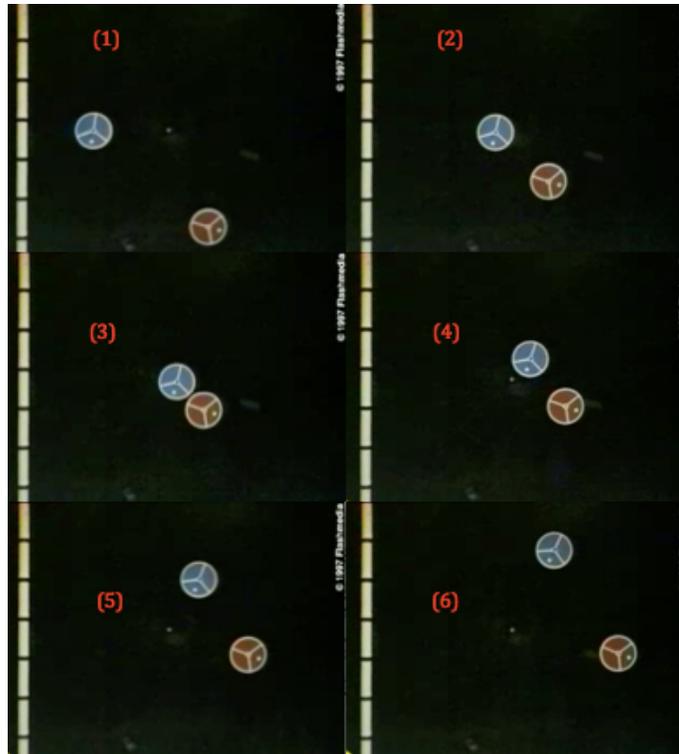

Figure 1: Six snapshots taken from the video [6]. The time ordering of the snapshots is given. The first two snapshots are from before the collision. Snapshot (3) is the frame of closest approach. Frames (4), (5), and (6) are after the collision.

## Solution of the Problem

Consider two pucks (of possibly different masses), which collide in 2D. We are given the initial velocities, $\mathbf{v}_{1,i}$ and $\mathbf{v}_{2,i}$, and are interested in predicting the final velocities, $\mathbf{v}_{1,f}$ and $\mathbf{v}_{2,f}$. During the brief moment the two pucks are in contact, we can define the plane of contact to be the plane tangent to either puck at the contact point. The normal to the plane of contact at the interaction point is the line of impact. (See Figure 2.)

We introduce a coordinate system such that the *x*-axis is aligned along the line of impact, and the *y*-axis is on the plane of impact (so that the motion is on the *x*-*y* plane), so that the impulses are along only one of the coordinate axes. The problem is analyzed in the laboratory frame (and not, say, in the center-of-mass frame) because the laboratory frame directly corresponds to the experiment done by the students. In our model the collision is frictionless, i.e., no tangential forces (friction) impart rotation. This model is equivalent to stating that there is zero tangential COR, or that no kinetic energy is converted to rotational kinetic energy. Kinetic energy is still not conserved in this model, as some energy is converted into the mechanical deformation of the colliding pucks, and is subsequently dissipated. This model is a justified approximation for the collision we analyze [6].

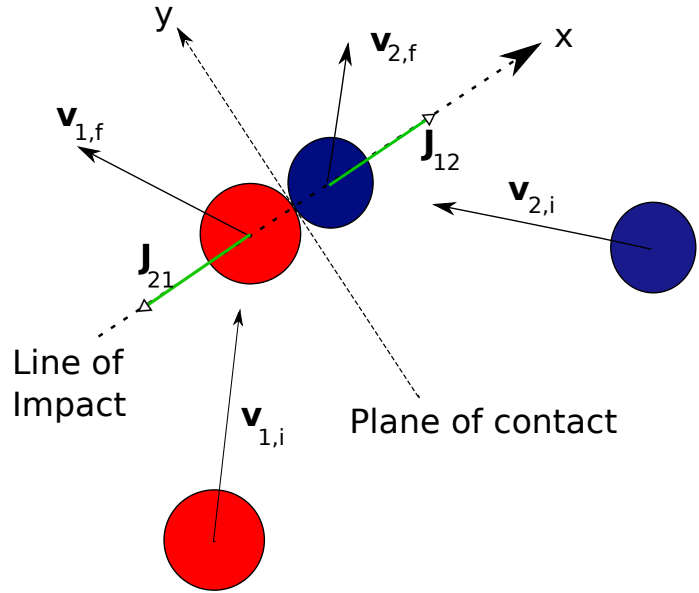

Figure 2: Puck 1 (red) collides with Puck 2 (blue). At the moment of collision the *plane of contact* is tangent to both pucks at the impact point. The *line of impact* is a normal to the plane of contact. The initial velocities are $\mathbf{v}_{1,i}$ and $\mathbf{v}_{2,i}$. The final velocities are $\mathbf{v}_{1,f}$ and $\mathbf{v}_{2,f}$. The impulse that puck 1 experiences because of puck 2, $\mathbf{J}_{21}$, and the impulse that puck 2 experiences because of puck 1, $\mathbf{J}_{12}$, are shown in green. Notice, that in our model the impulses are along the line of impact.

The key point to notice is that in this model the impulses are strictly along the line of impact (and result from normal forces). A component of the interaction along the plane of contact would require the two pucks to also roll or slide against each other, and then friction would impart rotation. It is also important that the line of impact goes through the centers of mass of both pucks. Otherwise, torques too will impart rotation. Our analysis is meant therefore as an approximation, valid when the initial and final rotational kinetic energies are negligibly small. This is the case for the collision [6].

The common approach is to write the condition for the conservation of the total momentum of the system, in addition to an energy condition. Our approach is as follows: The (normal) COR, *e*, is defined as

$$(1) \quad e = \frac{|v_{2,f,x} - v_{1,f,x}|}{|v_{2,i,x} - v_{1,i,x}|},$$

which is just the ratio of the final relative speed of the two pucks to its initial value along the line of impact. (The normal COR depends on the materials that the two pucks are made of, and depends only weakly on other parameters such as the impact speed, as long as the impact is not so violent as to permanently deform the pucks. The tangential COR, on the other hand, depends not just on bulk properties, but also on surface properties and also on the impact speed. This property of the tangential COR is our motivation to seek a problem wherein it can be ignored.) Generally, $0 \leq e \leq 1$, such that $e = 1$ for the perfectly elastic case, and $e = 0$ for the plastic case. (There are some cases for which *e* could assume values outside this range, which we do not consider here.) Stating $e = 1$ is equivalent to the statement of conservation of energy.

Next, consider the momenta of the two pucks:
$$(2a) \quad p_{1,i,x} + J_{21,x} = p_{1,f,x}$$
$$(2b) \quad p_{2,i,x} + J_{12,x} = p_{2,f,x},$$
where $\mathbf{J}_{ij}$ is the impulse that object $i$ imparts on object $j$. Adding these two equations we find
$$p_{1,i,x} + J_{21,x} + p_{2,i,x} + J_{12,x} = p_{1,f,x} + p_{2,f,x},$$
or
$$(3) \quad p_{1,i,x} + p_{2,i,x} = p_{1,f,x} + p_{2,f,x},$$
using $J_{21,x} = -J_{12,x}$, as is required by Newton's 3rd Law. Equation (3) is nothing but the $x$-component of the equation of conservation of momentum. It does not matter whether we use equation (3) or Eqs. (2a) and (2b) together with Newton's 3rd Law: In either case we have just one independent equation. (In the latter case we indeed add two equations, but also one dynamical constraint).

Next consider the $y$-direction: As there are no impulses in the $y$-direction, we find that
$$(4a) \quad p_{1,i,y} = p_{1,f,y}$$
$$(4b) \quad p_{2,i,y} = p_{2,f,y}.$$
We may, of course, add eqs. (4a) and (4b) to get the $y$-component of the equation for the conservation of momentum:
$$(5) \quad p_{1,i,y} + p_{2,i,y} = p_{1,f,y} + p_{2,f,y}.$$
But when we do so, we ignore the fact that in eqs. (4a) and (4b) there is more information than in eq. (5): It is not just the sum of the terms on either hand side of eq. (5) that is conserved: Each term is conserved separately! When we write eq. (5) we lose important information, which in practice reduces the number of equations that we have.

Let us now recount the equations and the unknowns: We have four unknowns: 2 components of the final momentum (or velocity) for each of the pucks, $p_{1,f,x}, p_{1,f,y}, p_{2,f,x}$, and $p_{2,f,y}$. But we also have four independent equations: Eq. (1) for the COR, Eq. (3) for the $x$-component of the equation for the conservation of momentum, and Eqs. (4a) and (4b) for the conservation of the $y$-component of the momentum for each puck separately.

### Application to the Collision Experiment

For the video collision we analyze [6] we typically give the students the (normal) COR. The students can also perform the experiments and take the videos themselves. Students can alternatively find experimentally the (normal) COR for a 1D collision, and then use it to determine the outcome of 2D collisions, as long as the systems is described by the simplified model.

The pucks in this video [6] are of equal mass, and their parameters ($m$ and $e$) are given in Table I, in addition to the (student-determined) initial velocities. Common programs such as Pasco Capstone or Tracker allow students to determine the velocities by tracking the motion of the pucks, and also allow students to rotate the coordinate axes and find the components of the velocities in the specified coordinate system. With these values the students first predict the final velocities $\mathbf{v}_{1,f}$ and $\mathbf{v}_{2,f}$ (including error propagation), and then measure them. The measured values too are included in Table I. Students find that their prediction agrees with the measurement to better than one standard deviation. They also find that momentum is conserved in this collision (with uncertainty smaller than one standard deviation), but that kinetic energy is not conserved. (The kinetic energy decreases by more than four standard deviations: The kinetic energies before and after the collision are also shown in Table I.)

| $m = 48g$ | $e = 0.56 \pm 0.03$ | **Initial velocity (m/s)** | **Final velocity (m/s)** |
|---|---|---|---|
| \multicolumn{2}{c|}{$\mathbf{v}_1$ **(Puck 1)**} | $-(0.41 \pm 0.02)\hat{\imath}$ $+ (0.30 \pm 0.02)\hat{\jmath}$ | $(0.25 \pm 0.02)\hat{\imath}$ $+ (0.32 \pm 0.02)\hat{\jmath}$ |
| \multicolumn{2}{c|}{$\mathbf{v}_2$ **(Puck 2)**} | $(0.43 \pm 0.02)\hat{\imath}$ $+ (0.47 \pm 0.02)\hat{\jmath}$ | $-(0.22 \pm 0.02)\hat{\imath}$ $+ (0.44 \pm 0.02)\hat{\jmath}$ |
| | | **Initial Kinetic Energy (J)** | **Final Kinetic Energy (J)** |
| \multicolumn{2}{c|}{**System**} | $0.0159 \pm 0.0010$ | $0.0098 \pm 0.0010$ |

Table I: The parameters of the colliding pucks and their measured initial and final velocities, and the system's initial and final values for the kinetic energy.

### Concluding remarks

The additional information that the interaction is only along the line of impact (and that the centers of mass are along the same) restore predictability: We no longer have an underdetermined set of equations, and we can find the full solution for the problem without having to provide, in addition to the system's initial conditions, also information about the final state.

As stated above, we prefer to present the problem by asking whether the conservation laws constrain the parameter space of the problem sufficiently much so that only a unique solution can be found. In 2D collisions that is not the case, and additional information on the system is needed. The information on the interaction that we give for the class of problems described herein allows us to do so. As a practical way to solve useful problems the approach of reducing the number of unknowns is effective. However, one needs to be careful to not confuse a practical method of resolving the underdetermined problem of the conservation laws with a necessary principle, which might have profound effect on causality and determinism in classical mechanics.

In practice, problems that involve the presented approach involve only a minor complication to the commonly available problems, and many problems can be posed at a level accessible to the students including the described laboratory exercise.

The author thanks Ted Forringer for making valuable comments on the manuscript. Early work on this paper was done while the author was at Georgia Gwinnett College

# Online Appendix

Textbooks we have surveyed approach this problem in three different ways: Type I textbooks ignore the problem, and in practice consider only particular cases in which there are only three unknown variables, such that one has just enough equations to solve the problem. In type Ia [7] only perfectly inelastic collisions (coefficient of restitution $e=0$) – or, equivalently, explosions – are discussed, so that the number of unknown variables reduces to three. Specifically, [7] only discusses an explosion of a single object in 2D, the time reversal of a perfectly inelastic collision. The end-of-chapter problems in [7] also discuss just the $e=0$ case. Another approach, Type Ib, is to ignore the problem, and make an assumption on a final condition [8]. When such an assumption is made, typically about the direction of one of the objects after the collision or the angle between the two final velocities, the number of unknown variables is again reduced to three. In practice, Type I approach is both practical and useful, as it allows the student to practice the vector properties of momentum in a simple context.

Type II textbooks explain the problem yet do not resolve it. They state a general solution cannot be found, and then in practice solve problems in which the number of unknown variables is reduced to three, as in type I textbooks. [9,10] For example, in [9] it is asked, "how can we solve for the unknowns?" and an answer is provided, "[t]he simple answer is, we can't!" but without an explanation of why, or what would be needed in order to solve for the unknowns. Ref. [10] makes even a stronger statement: "[W]e cannot determine the outcome of the collision without some additional information about the final velocities." Taken at face value, the latter statement is incorrect: As we shall see below, the outcome of 2D collisions can be determined even when information about the final velocities is not given (if other information on the detail of the interaction is given). But this statement might be rather confusing for students in addition to just being incorrect: The statement that (partial) final conditions of a system, in addition to initial conditions, are necessary in order to determine its future evolution challenges the determinism of classical mechanics and is anti-causal.

Type III textbooks explain the problem, hint at the correct resolution of the problem, but stop short from giving a full explanation. They still make an assumption on a final condition of the system in order to solve practical problems. [11]

If one objects to the statement regarding the required predictability of physical theory based on the probabilistic nature of quantum mechanics or on chaos theory, recall that also in quantum mechanics one still predicts the probability of the final state of the system, and that the Schrödinger equation is deterministic. In deterministic chaos theory the main issue is the extreme sensitivity of the system to initial conditions. But if the initial conditions were to be known exactly, the final state of the system would still be found uniquely. Be that as it may, here we are interested in a non-chaotic classical mechanics problem. It would be unsettling if a deterministic solution cannot be found!